\documentclass[oldversion]{aa}

\usepackage{epsfig}
\usepackage{graphics}
\usepackage{float}
\usepackage{amsmath}
\usepackage{multirow}
\usepackage{longtable}
\usepackage{rotate}
\usepackage{array}
\usepackage{subfigure}
\def\kms{\ifmmode{\rm km\,s^{-1}}\else\hbox{$\rm km\,s^{-1}$}\fi}

\setlongtables

\begin{document}

\title{An extensive grid of mass fluxes for Galactic O stars}

\author{L.B.Lucy}
\offprints{L.B.Lucy}
\institute{Astrophysics Group, Blackett Laboratory, Imperial College 
London, Prince Consort Road, London SW7 2AZ, UK}
\date{Received ; Accepted }

\abstract{A previously-described code for constructing moving reversing
layers (MRL) is improved by replacing a two-parameter model for 
$g^{\it l}(v)$, the radiative acceleration due to lines, with a flexible 
non-parametric description, thus allowing a greater degree of dynamical
consistency to be achieved in modelling turbulent transonic flow in the outer
atmospheric layers of O stars. With this new code, mass fluxes $J$
are computed at fifty-seven points in $(T_{\rm eff}, g)$-space. Specifically,  
$J$'s are computed for all Lanz-Hubeny (2003) NLTE atmospheres with
$T_{\rm eff}(kK) \in (27.5, 55)$ and $log \: g(cm \: s^{-2}) \leq 4.5$.
Differences with widely-used mass-loss formulae are emphasized, and 
opportunities for {\em differential} spectroscopic tests identified. 
}

\authorrunning{Lucy}
\titlerunning{Mass fluxes}
\maketitle

\section{Introduction}

In an earlier paper (Lucy 2007; L07), the moving reversing layer (MRL) theory
of Lucy \& Solomon (1970; LS70) was updated by incorporating an extensive line
list and improving the treatment of line formation.
In addition, the models were, in effect, grafted onto the TLUSTY static
NLTE O-star atmospheres of Lanz \& Hubeny (2003) by 
imposing the TLUSTY emergent continuum flux distribution as the
lower radiative boundary condition and
by matching ionization fractions at $T \approx 0.75 T_{\rm eff}$.

The motivation for reviving MRL theory was the conflict between 
observed and predicted mass loss rates ($\Phi$), which
had led several spectroscopic groups to question the theory of 
radiatively-driven winds (e.g., Bouret et al. 2005; Fullerton et al. 2006).
But, as emphasized in L07, the $\Phi$'s being tested were not 
obtained by
solving the equations governing the dynamics of radiatively-driven
winds but are the values $\Phi_{V}$  derived by Vink et al. (2000)
with a refined version of the semi-empirical Monte Carlo (MC) method of
Abbott \& Lucy (1985).
Logically, therefore, 
the conflict could arise from the specific assumptions of Vink et al.
rather than from a failure of the radiative-driving
mechanism.  

In L07, MRL models were used to explore 
the sensitivity of the eigenvalue $J = \Phi/ 4 \pi R^{2}$ to
$v_{t}$, the microturbulent velocity. Crucially, this parameter affects the
flux irradiating lines as matter is driven through the sonic point; and
an increase in $v_{t}$ from 10 to 15 km/s was found to decrease 
$J$ by $\approx 0.3$ dex.  
Thus, a physical effect was identified that, by reducing the predicted 
$\Phi$'s, might partially contribute to 
resolving the conflict.

More recently (Lucy 2010; L10), MRL theory was used to 
investigate individual stars. This was prompted by the work of Marcolino 
et al. (2009) on the {\em weak wind problem}, 
the major discrepancy for late-type O dwarfs between their $\Phi_{V}$'s
and observational estimates. To investigate this, a grid of 29
models was computed from which $J$'s for particular stars could be obtained
by interpolation. The results were encouraging: although the extremely
low and uncertain $J$'s estimated by Marcolino et al. were not matched,
the predicted
$J$'s were $\approx 1.4$ dex lower than the $J_{V}$'s. Moreover, when
$J$'s were interpolated for the two strong-wind O4 stars analysed by Bouret
et al.(2005), the results were consistent with the low values found when 
these authors took wind clumping into account.

In the above investigation, the L07 code was deliberately not
changed, thereby avoiding any suspicion that adjustments were motivated by
the observational data requiring explanation. But with the
technique's usefulness thus demonstrated, a possibly significant flaw 
is now addressed, namely the local departures from
dynamical consistency that result from the simple two-parameter 
representation of line driving. This further development is especially 
appropriate since dynamical consistency in modelling transonic flow was
identified in L10 as the key to accurate predictions of $J$ and $\Phi$. 
Accordingly, the primary purposes of this paper
are first to describe how such improved models can be constructed 
and then to compute $J$'s for all TLUSTY
atmospheres relevant for Galactic main-sequence O stars.

Throughout this paper $J$'s are in units $gm/s/cm^{2}$ and $\Phi$'s
in $\cal{M}_{\sun}$$/yr$.

\section{Improved solution technique}

As in previous papers, transonic flow is assumed to be stationary, isothermal,
and plane-parallel. The equation of motion can then be written as
\begin{equation}
    (v^{2}- a^{2}) \: \frac{1}{v} \frac{d v}{d x} = -g_{\rm eff}
\end{equation}
Here $a$ is the isothermal speed of sound, and $g_{\rm eff}= g - g_{e} - g^{\it l}$
is the effective gravity,   
where $g_{e} = \Gamma_{e} g$ and $g^{\sc l}$ are the radiative accelerations
due to electron- and line scatterings, respectively.

For given stellar parameters, we wish to find the solution of Eq. (1) 
such that the flow accelerates smoothly from sub- to supersonic velocities.
This is achieved by finding the particular mass flux $J$ that gives
$g_{\rm eff} = 0$ at the sonic point $v = a$, thus avoiding a singularity when
integrating Eq. (1). 

When approximated by a MC estimator, $g^{\it l}(v)$ is not analytic, and so
solving Eq. (1) and determining its eigenvalue $J$ is
not a conventional excercise in integrating an ODE. 
Accordingly, in L07, a two-parameter formula was adopted for $g^{\it l}(v)$ 
that
automatically gives $g_{\rm eff} = 0$ at $v = a$, thus allowing the
singularity-free stratification of the MRL to be obtained with a conventional
integration of Eq. (1). 
The MC transfer calculation was then carried out in this stratified medium, 
resulting in estimates
$\tilde{g}^{\it l}$ for each layer of the MRL. The challenge then was to
find the values of $J$ and of the parameters $\delta$ and $s$ that brought 
$\tilde{g}^{\it l}$ into optimal agreement with $g^{\it l}(v;J,\delta,s)$.

Because of this parametric approach, the solutions
obtained had noticeable residuals
$\Delta g^{\it l} = \tilde{g}^{\it l} - g^{\it l}$  - see Fig.3 in L10 - 
implying some uncertainty
in the predicted $J$'s. Although evidence was presented that $J$ is 
moderately insensitive to departures from detailed local dynamical
consistency, it is
nevertheless desirable to eliminate this weak point in MRL theory.

\subsection{Non-parametric $g^{\it l}(v)$}

To allow the $\tilde{g}^{\it l}$'s to be accurately modelled by 
$g^{\it l}(v)$, this function is constrained to pass through the discrete set
of points 
$(g^{\it l}_{i},v_{i})$, with $v_{1} < v_{2} < ...< v_{I}$. 
The required continuous function $g^{\it l}(v)$ is then constructed as follows: 
by linear logarithmic interpolation between neighbouring points for
$v \in (v_{1},v_{I})$; 
by setting $g^{\it l}(v) = g^{\it l}_{1}$ for $v < v_{1}$; and by 
extrapolating the power law from the interval $(I-1,I)$ for $v > v_{I}$.

The discrete representation extends from a small subsonic
velocity $v_{1}$ to a supersonic velocity $v_{I} \geq 2a$, with spacing chosen
to model the often sharply changing velocity gradient as the sonic point is
approached - see Fig.1 in L07. The $k$-th point is located at the
sonic point - i.e., $v_{k} = a$ - and the corresponding value of 
$g^{\it l}_{k}$ is constrained to be $g_{*} = g-g_{e}$, so that the regularity
condition is again automatically satisfied.

\subsection{Stratification}

With the $g^{\it l}(v)$ thus defined,
the MRL's stratification is obtained as described in 
Sect. 2.3 of L07, namely by two initial-value integrations of Eq. (1) starting
at $v = a$, one for
$v < a$ and one for $v > a$. 

To avoid a singularity at $v = a$, the initial
velocity gradient must be such that 
\begin{equation}
  \left( v \frac{dv}{dx} \right)_{a} = 
        \frac{1}{2}\: \left( \frac{d \: ln \: g^{\it l}}
                                      {d \: ln \: v} \right)_{a} g_{*}
\end{equation}
Now, if  $g^{\it l}(v)$ were an analytic function, the logarithmic derivative
in 
Eq. (2) would be the same for both inward and outward integrations. But
the adopted piecewise-linear segmented representation of $log \: g^{\it l}$ is
not analytic: although continuous, its derivative is in general discontinuous
at $v = v_{i}$ and thus may be so at $v_{k} = a$. Accordingly, for
the inward and outward integrations, the logarithmic derivatives are the slopes
of the $(k-1,k)$ and $(k,k+1)$ segments, respectively. 

The discontinuities in the derivatives of $g^{\it l}(v)$ allow the
representation
to approximate curvature in ($log \: g^{\it l} - log \: v$)- plots. Of course,
as $I \rightarrow \infty$,
unlimited accuracy can be achieved, and the discontinuities then
$\rightarrow 0$.

\subsection{An example}

Model $t400g375$, with parameters $T_{\rm eff} = 40,000K, log \: g = 3.75$ and
$v_{t} = 10 km/s$, illustrates the improved technique. 

Fig.1 shows the first steps in the search for $J$. The starting values
for $g^{\it l}_{i \neq k}$ are obtained from the two-parameter formula -
see Eq. (1) in L10 -
with $\delta = 0.5$ and $s = 1.5$ -i.e., a broken power law, with the
switch to a rapidly increasing  $g^{\it l}(v)$ occurring at Mach number
$m = v/a = 0.63$. With $g^{\it l}(v)$ thus fixed, several
models were computed with varying $J$ in order to locate the root of
$Q_{1,2}(J) = 0$ - see Sect.2.3 and Fig.2 in L10. The result, $J = -5.64$ dex,
is then such that, as matter accelerates from $m_{1} = 0.5$ to $m_{2} = 2.0$, 
the work done by the gradients of gas and radiation pressures
accounts for the gain in mechanical energy per gm.
Nevertheless, the non-vanishing residuals $\Delta g^{\it l}$ imply
that the $I-1$ values $g^{\it l}_{i \neq k}$ require adjustment. 

Ideally, corrections to $g^{\it l}_{i \neq k}$  should be derived from the 
$\Delta g^{\it l}$ with an algorithm analogous to the temperature-correction
procedures in stellar atmosphere theory. But here a trial-and-error procedure
is followed based on inspection of plots such as 
Fig.1. Thus, Fig.1 shows that a steeper slope than
$s = 1.5$ is required for $m > 1$ and that the $g^{\it l}_{i}$ should be
increased by $\approx 0.1$ dex for $m \in (0.1,0.4)$. 

With the  $g^{\it l}_{i \neq k}$ thus adjusted, a new sequence of models
is computed, the modified root of $Q_{1,2} = 0$ derived, 
and an updated version of Fig.1 plotted.
This iterative procedure is
continued until a satisfactory degree of convergence is achieved. In this
case, the final model has $J = -5.72$ dex and is plotted in Fig.2.
The iteratively-corrected function $g^{\it l}(v)$ now agrees closely with the
$\tilde{g}^{\it l}$'s. A dynamically consistent model of turbulent
transonic flow has therefore been constructed.

\begin{figure}
\vspace{8.2cm}
\includegraphics{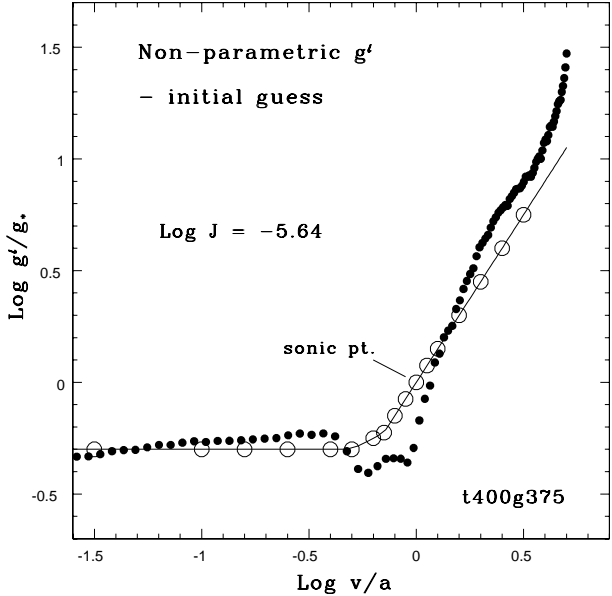}
\caption{Search for dynamical consistency. The open circles are 
the initial discrete representation
($g^{\it l}_{i}, v_{i}$) for model $t400g375$, and the connecting solid line
is the resulting continuous function $g^{\it l}(v)$.  
The MC estimates $\tilde{g}^{\it l}$ are plotted as filled circles and
correspond to the mass flux $J = -5.64$ dex that gives $Q_{1,2} = 0$. }
\end{figure}
\begin{figure}
\vspace{8.2cm}
\includegraphics{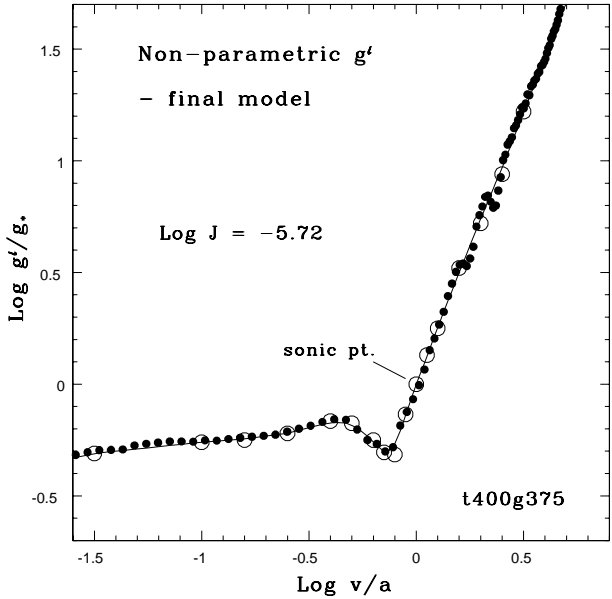}
\caption{Search for dynamical consistency. Symbols and model parameters as in
Fig.1. The
interatively-corrected representation ($g^{\it l}_{i}, v_{i}$) is shown 
together with the resulting MC estimates $\tilde{g}^{\it l}$  when
 $J = -5.72$ dex.}
\end{figure}

\section{Microturbulence}

As demonstrated in L07, the $J$'s predicted by MRL theory are sensitive
to $v_{t}$, which must therefore be included with $T_{\rm eff}$ and $g$ when
comparing with observational data. 

Given the importance of this sensitivity, the improved code is now 
applied to the model $t400g375$ 
in order to check and extend the analysis of L07. As in that investigation,
when $v_{t}$ is varied, the b-values and incident flux distribution are from
the TLUSTY model with $v_{t} = 10 km/s$. 

Solutions ranging from $v_{t} = 0$ 
 - i.e., pure thermal broadening in the lines' Doppler cores - 
to  $v_{t} = 20 km/s$, corresponding to near sonic turbulence, are plotted
in Fig. 3, together with the two solutions from L07 at $6.7$ and 
$10 km/s$. The sensitivity to  $v_{t}$ is confirmed.

\begin{figure}
\vspace{8.2cm}
\includegraphics{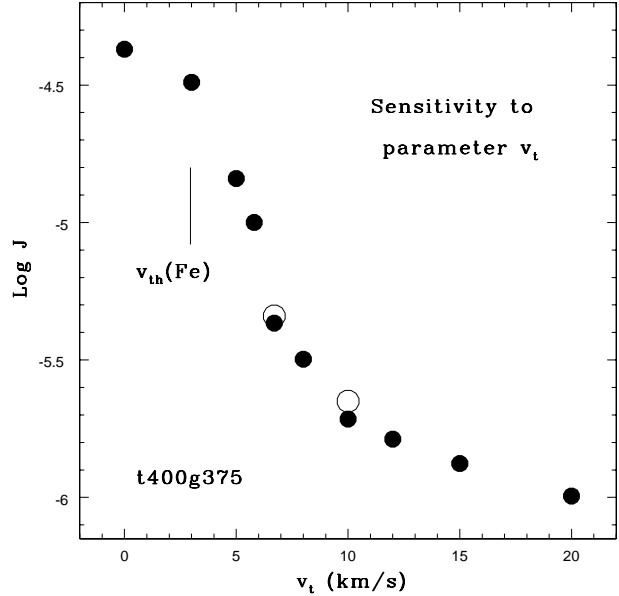}
\caption{Sensitivity of the eigenvalue $J$ to microturbulent velocity $v_{t}$
for model $t400g375$. Filled circles are solutions obtained as described in
Sect. 2; open circles are solutions from L07.  
The thermal speed of Fe ions is indicated.}
\end{figure}

From the new results between $6.7$ and $15 km/s$, the logarithmic
slope at $10 km/s$ is
\begin{equation}
              \frac{\partial \: log J  }{\partial \: log v_{t}} = -1.46
\end{equation}
This value is recommended for propagation-of-error calculations of
$\sigma_{log J}$ for stars with $v_{t}$'s
comparable to the canonical $v_{t} = 10 km/s$.

Fig.3 shows that, in principle, the parameter $v_{t}$ could bring about
differences in $J$ by $\ga 1$ dex at fixed $T_{\rm eff}, g$. But as this seems not 
to happen for real stars, the mechanism exciting and maintaining 
microturbulence is presumably preventing the independent variation of this
parameter, resulting perhaps in a functional dependence of $v_{t}$ on 
$T_{\rm eff}, g$.

In view of the strong damping expected for turbulence with $v_{t}/a \sim 0.5$,
work done by radiative forces is almost certainly required for its 
maintenance, in 
which case the phenomenon of microturbulence is not unrelated to radiatively-
driven outflows. In fact, Fig.3 suggests a direct, causal relationship as
follows: Given the observational evidence that wind-clumping occurs
shortly after the sonic point (Bouret et al. 2005), some clumps may well lose
their net outward driving and thus fall back into the photosphere 
(e.g., Howk et al. 2000)
where their dissipated kinetic energy could excite and maintain local
turbulence.
If the fall-back fraction decreases with decreasing $J$, a feedback loop
operates, so that for a given star only one pairing $(J,v_{t})$ is possible.

This conjectured coupling of  $J$ and $v_{t}$ is supported by the difficulty of
obtaining satisfactory high mass flux MRL's when $v_{t} \rightarrow 0$. 
The $J$'s for $v_{t} \la 5 km/s$ in Fig. 3 are estimated by achieving dynamical
consistency only for 
$v/a < 1$ since the extension to $v/a \ga 2$ is not possible because $g_{\rm eff}$
becomes positive - see Fig.4. Admittedly, this breakdown of the iterative 
procedure is 
code-specific, resulting from choosing $v$ rather than height $x$ as the
independent variable. Nevertheless, even if the code were
reconfigured to allow dynamical consistency to be extended to supersonic
velocities, the resulting
non-monotonic velocity law implies a density inversion at $v/a \sim 2-3$.
Such a stratification is surely unstable    
and thus a possible origin of clumping and of infalling blobs.

\begin{figure}
\vspace{8.2cm}
\includegraphics{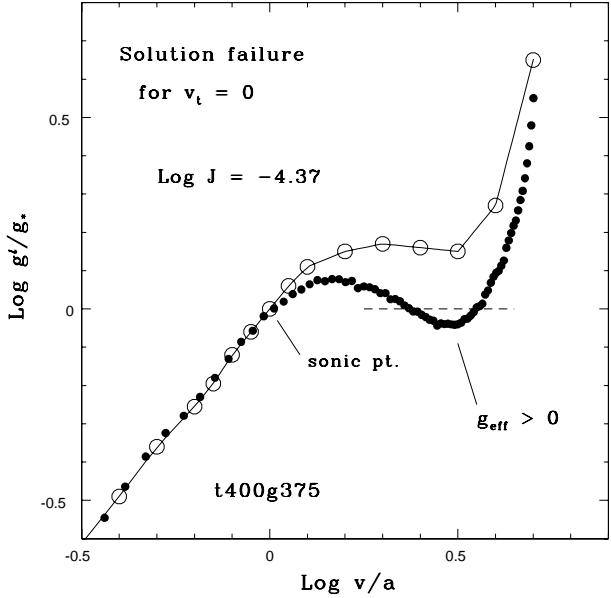}
\caption{Failed search for dynamical consistency when $v_{t} = 0$
for model $t400g375$. Symbols as in Fig.1.}
\end{figure}

\subsection{A spectroscopic test}  

This sensitivity of $J$ to $v_{t}$ is in stark contrast
to CAK theory (Castor, Abbott\& Klein 1975), which posits
that the properties of 
radiatively-driven winds can be derived on the basis of the Sobolev
approximation. In this approximation, $g^{\it l}(v)$ is independent
of the lines' absorption profiles and therefore independent of $v_{t}$'s
contribution to the width of the Doppler core.

In principle, this can be tested observationally, and is best done
{\em differentially}. Thus stars differing in  $v_{t}$ but not widely
separated in $(T_{\rm eff},g)$-space should be subjected to identical observing and
diagnostic procedures to see if $\Delta \Phi$ can be understood without (CAK),
or only with (MRL), a contribition from $\Delta v_{t}$.

\section{Computed mass fluxes}

In this section, the improved technique of Sect.2 is used
to recalculate the 29 models in Table 1 of L10 and then to add a further
28 models
in order to provide a rather complete coverage of $(T_{\rm eff},g)$-space
for H-burning O stars. As before, the models'
composition is solar with $N_{He}/N_{H} = 0.1$ (Grevess \& Sauval 1998),the
included metal ions are as in Table 1 of Lanz \& Hubeny (2003), and
$v_{t} = 10 km/s$.

\subsection{O-star grid}

Mass fluxes $J$ for 57 models are given in Table 1. The grid is determined by
the availability of TLUSTY atmospheres (Lanz \& Hubeny 2003) and is
complete for their models with $log \: g \leq 4.5$. Because O stars on the ZAMS
have $log \: g \approx 4.2$ - see Fig.4 in L10, $J$'s have not been computed
for $log \: g = 4.75$.

The coverage provided by Table 1 allows $J$'s to be
determined for all Galactic O stars by interpolation - or a slight
extrapolation in a few cases. If, in addition to
$T_{\rm eff}, g$ and $v_{t}$, a star's distance is known, its
radius can be computed and therefore also $\Phi = 4 \pi R^{2}J $. Given the
detailed diagnostic modelling of numerous O stars in recent years, a critical
evaluation of MRL theory may be possible with existing data. This
is not attempted here and is,
in any case, best carried out by investigators familiar
with the uncertainties of analysing circumstellar spectra.  

The data in Table 1 is also relevant for investigations of stellar evolution
with mass loss, for computing the latitude dependence of mass loss for rapidly-
rotating stars, and for calculating its radial dependence 
for accretion disks.

The dependence of $J(T_{\rm eff},g;v_{t})$ on $T_{\rm eff}$ and $g$ is shown in
Figs. 5 and 6. The most striking feature is the departure from
the expected monotonic increase of $J$ with increasing $T_{\rm eff}$
that occurs when $T_{\rm eff} \la 30,000K$ and $log g \ga 3.9$. As discussed in L10,
this prediction
of MRL theory offers at least a partial explanation of the {\em weak-wind
phenomenon}. Interestingly, the minimum at 
$T_{\rm eff} \approx 30,000K$ is deeper for $g > g_{ZAMS}$ 

\begin{figure}
\vspace{8.2cm}
\includegraphics{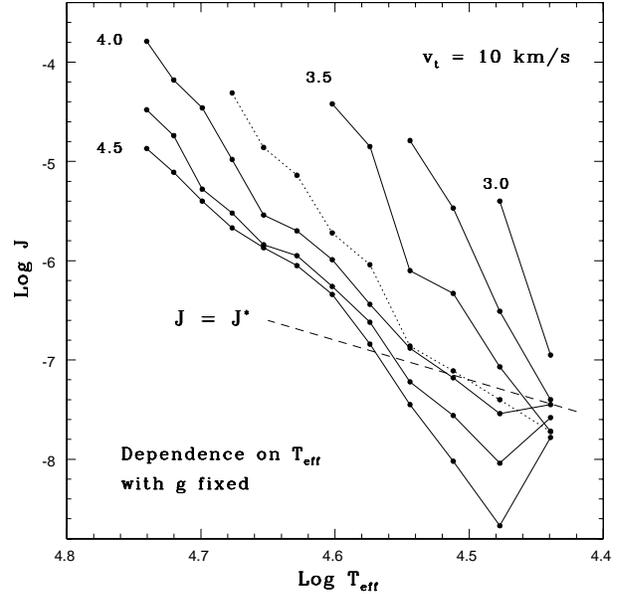}
\caption{Mass flux $J$ as a function of $T_{\rm eff}$ for 
$log \: g = 3.00(0.25)4.50$. The data are from Table 1. The dashed line
$J = J^{*}$ defines the boundary of the weak-wind domain - see L10.}
\end{figure}
\begin{figure}
\vspace{8.2cm}
\includegraphics{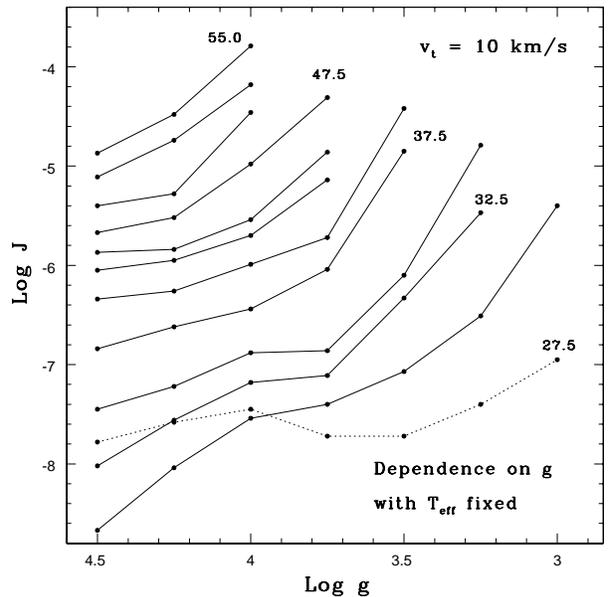}
\caption{Mass flux $J$ as a function of $g$ for 
$T_{\rm eff} = 27.5(2.5)55.0 \: kK$. The data are from Table 1.}
\end{figure}

\begin{table}

\caption{Computed mass fluxes $J$.}

\label{table:1}

\centering

\begin{tabular}{c c c c c}

\hline\hline

$log \: g$    & $T_{\rm eff}(kK)$ & $log J$ & \; \; \;  $T_{\rm eff}(kK)$ &  $log J$ \\

\hline
\hline

 4.50  &  27.5   &  -7.78   &   \; \; \;   42.5  &  -6.05  \\

       &  30.0   &  -8.67   &   \; \; \;   45.0  &  -5.87  \\

       &  32.5   &  -8.02   &   \; \; \;   47.5  &  -5.67  \\

       &  35.0   &  -7.45   &   \; \; \;   50.0  &  -5.40  \\

       &  37.5   &  -6.84   &   \; \; \;   52.5  &  -5.11  \\

       &  40.0   &  -6.34   &   \; \; \;   55.0  &  -4.87  \\

\cline{1-5}

 4.25  &  27.5   &  -7.58   &   \; \; \;   42.5  &  -5.95  \\

       &  30.0   &  -8.04   &   \; \; \;   45.0  &  -5.84  \\

       &  32.5   &  -7.56   &   \; \; \;   47.5  &  -5.52  \\

       &  35.0   &  -7.22   &   \; \; \;   50.0  &  -5.28  \\

       &  37.5   &  -6.62   &   \; \; \;   52.5  &  -4.74  \\

       &  40.0   &  -6.26   &   \; \; \;   55.0  &  -4.48  \\

\cline{1-5}

 4.00  &  27.5   &  -7.45   &   \; \; \;   42.5  &  -5.70  \\

       &  30.0   &  -7.54   &   \; \; \;   45.0  &  -5.54  \\

       &  32.5   &  -7.18   &   \; \; \;   47.5  &  -4.98  \\

       &  35.0   &  -6.88   &   \; \; \;   50.0  &  -4.46  \\

       &  37.5   &  -6.44   &   \; \; \;   52.5  &  -4.18  \\

       &  40.0   &  -5.99   &   \; \; \;   55.0  &  -3.79  \\

\cline{1-5}

 3.75  &  27.5   &  -7.72   &   \; \; \;   40.0  &  -5.72  \\

       &  30.0   &  -7.40   &   \; \; \;   42.5  &  -5.14  \\

       &  32.5   &  -7.11   &   \; \; \;   45.0  &  -4.86  \\

       &  35.0   &  -6.86   &   \; \; \;   47.5  &  -4.31  \\

       &  37.5   &  -6.04   &   \; \; \;         &         \\

\cline{1-5}

 3.50  &  27.5   &  -7.72   &   \; \; \;   35.0  &  -6.10  \\

       &  30.0   &  -7.07   &   \; \; \;   37.5  &  -4.85  \\

       &  32.5   &  -6.33   &   \; \; \;   40.0  &  -4.42  \\

\cline{1-5}

 3.25  &  27.5   &  -7.40   &   \; \; \;   32.5  &  -5.47  \\

       &  30.0   &  -6.51   &   \; \; \;   35.0  &  -4.79  \\

\cline{1-5}

 3.00  &  27.5   &  -6.95   &   \; \; \;   30.0  &  -5.40  \\

\hline
\hline

\end{tabular}

\end{table}

\subsection{Accuracy}

The $J$'s in Table 1 have several sources of uncertainty.
The first originates from MC sampling errors - see Appendix A.  
The $\tilde{g}^{\it l}$'s sampling errors propagate via $Q_{1,2}$ 
into errors in $J$ when this quantity is
determined by locating the intercept $Q_{1,2}(J) = 0$ -
see Fig.2 in L10. Thus, for the final $t400g375$ model in Fig.2, the 
least squares solution is $log \: J = -5.715 \pm 0.014$. This estimate
of $\sigma_{log J}$ shows that MC noise is inconsequential since other errors
are surely far greater. 

The uncertainty $\sigma_{log J} \rightarrow 0$ as the number 
of MC quanta $N \rightarrow \infty$, but $J$ would still be  
subject to error because of residual departures from exact dynamical 
consistency if the continuous function $g^{\it l}(v)$ 
derives from only a {\em finite} number of points $(g^{\it l}_{i},v_{i})$.  

This second source of uncertainty can be estimated as follows:
Given that the non-parametric representation is a marked improvement over the
previous model, the average $|\Delta log J|$ between the models treated both
here and in L10 is a good estimate of the typical error of the previous
$J$'s and, at the same time, a conservative error estimate for the $J$'s in 
Table 1. The 29 $J$'s 
differ on average by only 0.10 dex, with the largest difference being 0.33 dex.

A third source of uncertainty is the cumulative effect of errors
in abundances, input physics and line-formation theory, many of which propagate
from the TLUSTY models.
A plausible guess is that these errors should rarely exceed 0.2 dex.

A fourth source of uncertainty is the derivation of $J$ from a plane-parallel
treatment of transonic flow, with back-scattering from $v > 5a$ neglected.
This is investigated in Appendix B and found to well-justified.

A fifth and probably dominant source of uncertainty is the sensitivity to 
the throttling effect of turbulent line broadening (L07 and Sect.3 above).  
Given our ignorance as to the source and nature of photospheric
turbulence, this has perforce been investigated in the microturbulent limit,
and strong sensitivity is found. Thus, from Fig.3, we see that,
with $v_{t} = 10 km/s$, photospheric turbulence reduces $J$ by $\approx$ 1.3 dex
from its value for laminar flow - i.e., pure thermal broadening. Within
the context of the microturbulent model, an error of $\pm 2 km/s$ at
$v_{t} \approx 10km/sec$ implies $\sigma_{log J} \approx 0.13$ according to
Eq. (3).

Given these uncertainties, tests of MRL theory, either 
spectroscopically or via stellar evolution calculations, might reasonably
allow for an uncertainty of $\pm 0.2$ dex in the $J$'s given in Table 1. But
if a test reveals systematic discrepancies $> 0.4$ dex, a contradiction can
be claimed.

\section{Comparisons with mass-loss formulae}

In this section, the MRL mass fluxes are compared to widely-used
mass-loss formulae. The aim here is not a comprehensive discussion
but to call attention to opportunities for {\em differential} testing, as already
discussed in Sect. 3.1.

\subsection{CAK}
 
In their recent discussion of WNH stars, Smith \& Conti (2008) sketch an
evolutionary scenario based on an O-star mass-loss formula extracted from  
CAK theory. For fixed $T_{\rm eff}$, their formula gives
$J$'s dependence on $g = g_{e}/\Gamma_{e}$ as 
\begin{equation}
  J = J_{0} \; \frac{\Gamma_{e}}{1 - \Gamma_{e}}
\end{equation}
where $J_{0}$ is the mass flux when 
$\Gamma_{e} = 0.5$. Eq. (4) predicts that $J \rightarrow \infty$ as
$\Gamma_{e} \rightarrow  1$ - i.e., as the Eddington limit is approached
- and this is basic to their claim that a feedback 
process results in runaway mass loss late in the core-H burning phase   
of massive O stars.

\begin{figure}
\vspace{8.2cm}
\includegraphics{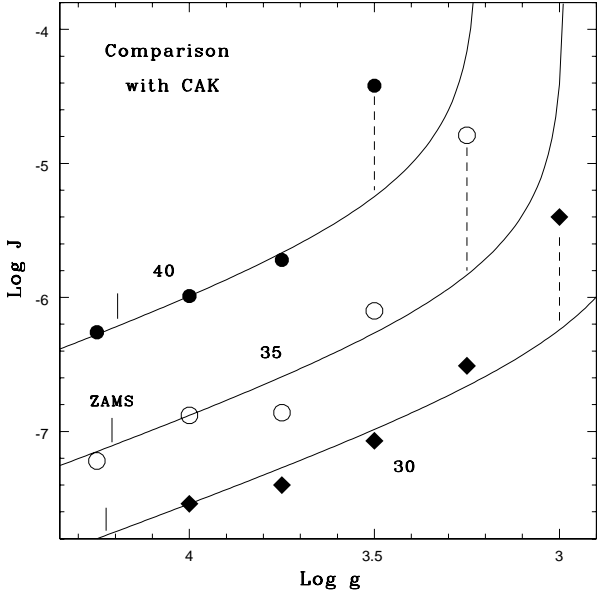}
\caption{Comparison of MRL mass fluxes $J$ with scaled CAK formulae for 
$T_{\rm eff} = 30, 35 $ and $40 kK$. The dashed vertical lines are the MRL-CAK
offsets discussed in the text.}
\end{figure}

In Fig.7, the behaviour of $J_{L}$ as $g \rightarrow  g_{e}$ is compared
to Eq.(4) when $J_{0}$ is chosen to match $J_{L}$
at $log \: g = 4$.
This comparison is carried out for
$T_{\rm eff} = 30,35$ and $40 kK$, for which the lowest $g$ TLUSTY models have
$\Gamma_{e} = 0.52, 0.54$ and $0.52$, respectively. Remarkably, MRL theory
predicts greatly enhanced mass loss when a star is still well-detatched
from its Eddington limit. In each of the plotted sequences, the lowest $g$
model is offset from the scaled CAK formula by $\Delta J \sim 1$ dex.

If the Smith-Conti scenario were supported by actual stellar evolution tracks
and accurately fitted several well-observed WNH stars, then MRL theory could
be immediately dismissed as over-predicting $J$ for evolved O stars. But neither
of these circumstances obtains, and this enhanced mass loss already
at $\Gamma_{e} \approx 0.5$ will likely prove favourable to their scenario. 
Clearly, further work is required on the evolution of mass-losing massive stars.

In addition to its relevance for evolutionary scenarios, Fig. 7 suggests
another {\em differential} test to distinguish MRL and CAK theories.       
In this case, pairs of stars with closely similar $T_{\rm eff}$'s but
markedly different $g$'s should be observed and analysed identically to
see which $g$-dependence in Fig. 7 is favoured.

\subsection{Vink et al.}

A model from Table 1 cannot be directly compared to the Vink et al. formula
since the latter requires {\em three} fundamental stellar parameters {\em and}
$v_{\infty}$. Accordingly, the comparison is carried out for the ZAMS models
of Pols et al. (1998) with $Z=0.02$. Each point on the ZAMS gives 
$\cal{M}$,$R$ and $L$ which, with the additional assumption that  
$v_{\infty}/v_{esc} = 2.6$ (Lamers et al. 1995), allows $\Phi_{V}$ to be
computed from Eq. (12) of Vink et al. (2000). This can then be compared to
$\Phi_{L} = 4 \pi R^{2} \times J(T_{\rm eff},g)$. Here $J$ is obtained by
simple bivariate interpolation (Abramowitz \& Stegun 1965) from the 
four surrounding entries in Table 1,
taking the independent variables to be $log \: T_{\rm eff}$ and $log \: g$.

The two predictions for the ZAMS are plotted in Fig. 8. Throughout the entire
range $\Delta \Phi = \Phi_{V} - \Phi_{L} > 0$. At $T_{\rm eff} =27,650 K$, the
offset is 0.40 dex, and this increases steeply to a maximum of 1.24 dex at 
$30,150 K$. Thereafter, $\Delta \Phi$ decreases - not quite 
monotonically - to reach a barely significant 0.27 dex at $T_{\rm eff} = 50,300 K$.

The huge difference at $T_{\rm eff} \approx 30,000K$ allows MRL theory to
{\em partly} explain the {\em weak wind phenomenon}, which arose when the
Vink et al. predictions were compared to mass loss estimates 
for late-type O dwarfs. In
fact, the relevant diagnostic analyses (Marcolino et al. 2009) appear still to
require $\Phi_{L}$ to be reduced by  $\approx$ 0.8 dex - see L10.

According to Fig.8, MRL theory predicts that a massive star's initial $\Phi$
is markedly less than $\Phi_{V}$. However, when its expanding
radius has reduced $g$ by $\approx 1$ dex, the situation reverses- see Fig. 7.
Thus, for example, if a mass-losing star reaches the point $(40,000K, 3.5)$ in
$(T_{\rm eff}, log\:g)$-space with $\cal{M/M_{\sun}}$$ = 80$, then $\Phi_{V} = -4.75$
dex, but $\Phi_{L} = -4.59$ dex, a factor 1.4 larger. 
To put this in context, the $120\cal{M_{\sun}}$ track of Limongi \&
Chieffi (2006) computed with Vink et al. mass loss has $log\:g = 3.56$ and 
$3.31$ with mass $93.6$ and $56.6 \cal{M_{\sun}}$, respectively, at its two 
crossings of $T_{\rm eff} = 40,000K$ during core H-burning.  

From the trends evident in Fig. 7, the MRL mass-loss enhancement
will be even greater for $g$'s smaller than given in Table 1.
Moreover, the Limongi-Chieffi track suggests that this part of parameter space
may well be accessed by real stars. Accordingly, Table 1 needs to be extended
to lower $g$'s.

\begin{figure}
\vspace{8.2cm}
\includegraphics{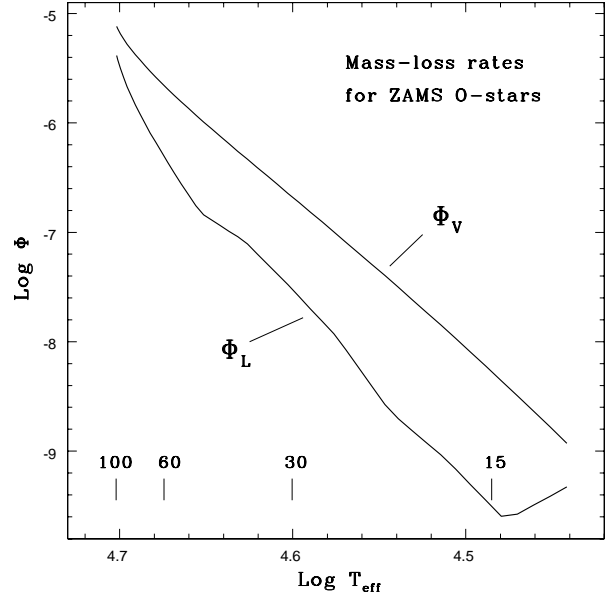}
\caption{Mass-loss rates for O-stars on the zero-age main sequence (ZAMS)
as a function of $T_{\rm eff}$, with masses in solar units indicated.
Predictions of MRL theory $(\Phi_{L})$ are plotted as well as values 
$(\Phi_{V})$ derived with the formula of Vink et al. (2000).}
\end{figure}

\section{Conclusion}

Motivated by the (partial) successes of MRL theory in reproducing
the reduced $\Phi$'s found by spectroscopists for O stars when wind-clumping 
is taken into account, the aim of this paper has been to
complete the coverage of the relevant $(T_{\rm eff},g)$-domain provided
by the TLUSTY atmospheres. To this end, the MRL code has been improved by 
adopting
a non-parametric description of $g^{\it l}(v)$, which has allowed a greater
degree of dynamical consistency to be achieved for turbulent transonic flow, and
thus more accurate $J$'s. The results of this effort are the 57
values of $log \: J$ in Table 1.

Interpolation in Table 1 allows $J$ to be derived for any O star with
measured $T_{\rm eff}$ and $g$. Moreover, a propagation-of-error calculation
gives $\sigma_{log J}$ if the standard errors of these two parameters and of
$v_{t}$ have been determined. This can be done for stars individually
and independently. However, given the difficulties of diagnostic analyses,
with the resulting possibility of systematic errors, there is merit
in performing {\em differential} tests as suggested in Sects.3.1 and 5.1.
If $v_{t}$ is a slowly-varying function of $T_{\rm eff}$ and $g$, the first
test may be indecisive. But the proposed test between the MRL
and CAK theories suggested by Fig.7 is feasible. Moreover, this test is
fundamental for stellar wind theory since it directly concerns the question:
where in the outflow is the mass-loss rate determined?

With regard to this question, the partial success of MRL theory in explaining
the {\em weak wind phenomenon} supports the LS70 argument that $J$ and therefore
$\Phi$ is determined by the regularity condition at the {\em sonic point}. This
would be decisive if the observed $\Phi's$ of the Marcolino et al. (2009) stars
were convincingly revised into agreement with the predictions of MRL theory. 
But since only weak C IV absorption is observed, there is little or no 
observational basis for such improved estimates. Other tests should therefore
be carried out.

\acknowledgement

I am grateful to M.Limongi and A.Chieffi for unpublished details of their
evolutionary tracks and to the referee, A. de Koter, for his detailed comments on
the proposed differential tests.

\appendix

\section{Precision of the estimator $\tilde{g}^{\it l}$}

The $\tilde{g}^{\it l}$ are derived from a MC simulation 
using estimator $A$, the summation over
pathlengths given in Eq. (10) of L07. This is expected to be more accurate
than estimator $B$, the summation of momentum tansfers from energy packets to
matter at the discrete line-scattering events. 
 
A test of the accuracy and convergence of $A$ and $B$ has been carried out
for model $t500g400$. In this test, the exact $g^{\it l}$ is taken to be
$\tilde{g}^{\it l}$ given by $A$ when $N$, the number of MC quanta, is 
$512 \times 10^{6}$.
Given this 'exact' $g^{\it l}$, the fractional errors of $\tilde{g}^{\it l}$
can be computed at smaller $N$ for $A$ and at all $N$ for $B$.
    
The mean absolute fractional errors for $0.5 < v/a < 2$ are plotted against
$N$ in Fig. A.1. As expected, errors for both $A$ and $B$ are 
$\propto 1/\sqrt{N}$. Also as expected, $A$ is the more accurate.
From the plotted least squares fits, 
$\epsilon_{A} = 0.488 \times \epsilon_{B}$. Thus, to achieve the same accuracy,
$B$ would have required $N$ to be increased by the factor 4.2. The 
saving of computer time with $A$ was essential in carrying out
the huge modelling effort required to produce Table 1.

A typical simulation has $N= 40 \times 10^{6}$ and so, according
to Fig. A.1., the typical error of $\tilde{g}^{\it l}$ with $A$ is
0.016 dex. But this is specific to model $t500g400$. At cooler $T_{\rm eff}$'s,
an increasing fraction of packets propagate through the MRL without undergoing
line scatterings, so that, for fixed $N$, the sampling error of 
$\tilde{g}^{\it l}$ increases. Thus, for model $t325g400$, the above error
increases to 0.20 dex.

\begin{figure}
\vspace{8.2cm}
\includegraphics{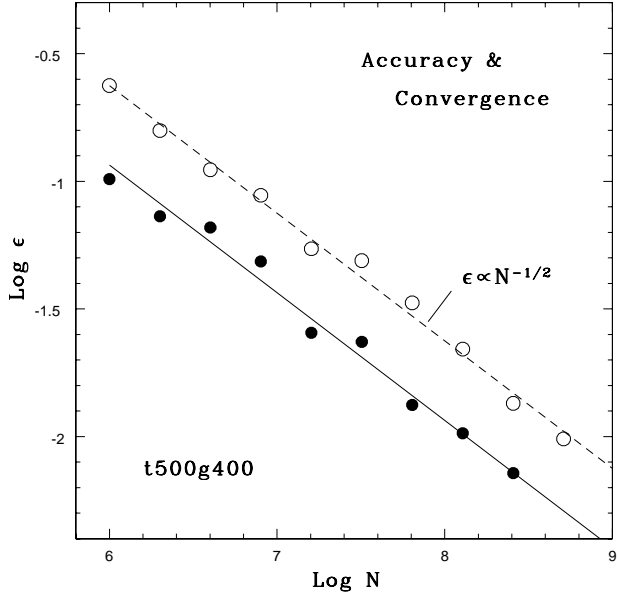}
\caption{Convergence and accuracy test of MC estimators. For model t500g400,
the mean absolute
fractional error $\epsilon$ of $\tilde{g}^{\it l} $ for $0.5 < v/a < 2$
is plotted against $N$, the number of energy packets.
The filled and open circles refer to estimators $A$ and $B$, repectively. }
\end{figure}

\section{Reflection probability}

When a MC quantum crosses the upper boundary of the MRL at height $x_{1}$ where
$v/a = 5$, it permanently exits the computational domain. Thus,
its dimensionless reflection probability $p_{1} = 0$, an assumption 
justified if the true $p_{1} \ll 1$.

An estimate of $p_{1}$ can be derived following the method of 
Abbott \& Lucy (1985; Sect III). We assume that radiative driving
dominates other mechanisms beyond $x_{1}$ and accelerates the wind to terminal
velocity $v_{\infty}$. On this assumption, increases in the mechanical 
luminosity
\begin{equation}
  {\cal L} (r) = \Phi \; (\: \frac{1}{2} v^{2} - \frac{G {\cal M} }{r} )
\end{equation}
are accounted for by a matching decrease in the radiative luminosity 
$L(r)$.
Thus, in the spherical shell $(r,r+dr)$, the $O(v/c)$ difference between
the rates at which matter absorbs
$d {\cal A}$ and emits $d {\cal E}$ radiant energy is
\begin{equation}
  d {\cal A} - d {\cal E} = \: d {\cal L} \: = \Phi \: (v \frac{dv}{dr} +g)\: dr
\end{equation}

If we now assume that the electron- and line-scatterings responsible for the
energy transfer absorb from a {\em radially-streaming} radiation field and emit
isotropically, then

\begin{equation}
  d {\cal E} = \frac{c}{v} \: d {\cal L} 
\end{equation}
Since this energy is radiated isotropically, the fraction propagating back
through $r = r_{1}$ - i.e., back into the MRL - is 
\begin{equation}
   w_{1} = \frac{1}{2} \; [\: 1- \sqrt{1-(z/z_{1})^{2}} \; ] 
\end{equation}
provided that no interactions intervene. Here 
$z=R/r$, where $R$ is the photospheric radius. 

Combining the above, we find that the luminosity of inwardly-propagating
radiation at $r_{1}$ is  
\begin{equation}
   L^{-}_{1} = \int_{r_{1}}^{\infty} w_{1} \: \frac{c}{v}  \: 
                               \frac {d {\cal L}}{d r} \: dr  
\end{equation}
The reflection probability is then $p_{1} = L^{-}_{1}/(L_{*}+L^{-}_{1})$, where 
$L_{*}$ is the luminosity the wind-free star. 

In evaluating $p_{1}$, we assume
$L^{-}_{1} \ll L_{*}$ and that the supersonic wind obeys a $\beta$-velocity
law with $\beta = 1$. The result is 
\begin{equation}
   p_{1} = \frac{\Phi}{\; \Phi_{\dagger}} \int_{0}^{z_{1}} w_{1}  \:
 [\: 1+\frac{\eta}{1-z}] \: dz  
\end{equation}
Here  $\Phi_{\dagger} = L_{*}/c v_{\infty}$ (Cassinelli \& Castor 1973)
and $\eta = (v_{esc}/v_{\infty})^{2}$,
where $v_{esc}$ is the escape velocity from $r = R$.

Values of $p_{1}$ have been computed for the ZAMS models in Fig.8. Thus, at 
$ {\cal M} =$ $30  {\cal M}_{\sun}$ with $v_{\infty}/v_{esc} = 2.6$, the ratio
$\Phi_{L}/\Phi_{\dagger} = 0.044$. Then, for $v_{1} = 5a$,
$z_{1} = 0.969$, and we find $p_{1} = 0.0071$. The neglect of back-scattering 
from the exterior wind is therefore justified. Moreover, since 
$1-z_{1} \ll 1$, the assumption of plane-parallel geometry is also justified.

But for very massive stars, the assumption $p_{1} = 0$ is less valid. 
Thus, for $ {\cal M} =$ $80  {\cal M}_{\sun}$, 
$\Phi_{L}/\Phi_{\dagger} = 0.384$, $z_{1} = 0.972$, and we find that
$p_{1} = 0.064$. Accordingly, when theory and observation agree to 
$\la 0.1$ dex, further progress will require an improved treatment of
transonic flow.

\end{document}